\documentclass[a4paper,fleqn]{cas-sc}
\usepackage{soul}
\usepackage[sort&compress,numbers]{natbib}
\usepackage{lineno}
\usepackage{setspace}
\usepackage{multirow}

\usepackage{cuted}

\def\tsc#1{\csdef{#1}{\textsc{\lowercase{#1}}\xspace}}
\tsc{WGM}
\tsc{QE}
\tsc{EP}
\tsc{PMS}
\tsc{BEC}
\tsc{DE}

\begin{document}
\let\WriteBookmarks\relax
\def\floatpagepagefraction{1}
\def\textpagefraction{.001}
\shorttitle{Intrinsic Properties of Bipolarons in Armchair Graphene Nanoribbons}
\shortauthors{Silva \textit{et~al}.}

\title [mode = title]{Intrinsic Properties of Bipolarons in Armchair Graphene Nanoribbons}

\author[1]{Gesiel G. Silva}
\author[2]{Wiliam F. da Cunha}
\author[2]{Marcelo L. Pereira J\'unior}
\author[2]{Luiz F. Roncaratti J\'unior}
\author[2,3]{Luiz A. Ribeiro J\'unior}
\cormark[1]
\ead{ribeirojr@unb.br}

\address[1]{Federal Institute of Education, Science and Technology of G\'oias, Luzi\^ania, G\'oias, Brazil}
\address[2]{Institute of Physics, University of Bras\'ilia, 70910-900, Bras\'ilia, Brazil}
\address[3]{PPGCIMA, Campus Planaltina, University of Bras\'{i}lia, 73345-010, Bras\'{i}lia, Brazil}

\cortext[cor1]{Corresponding author}

\begin{abstract}
We performed an investigation concerning bipolaron dynamics in armchair graphene nanoribbons (AGNRs) under the influence of different electric field and electron-phonon coupling regimes. By studying the response to the electric excitation, we were able to determine the effective mass and terminal velocity of this quasiparticle in AGNRs. Remarkably, bipolarons in narrower AGNRs move as fast as the ones in conjugated polymers. Our findings pave the way to enhance the understanding of the behavior of charge carriers in graphene nanoribbons.
\end{abstract}




\maketitle
\doublespacing

\section{Introduction}
\label{sec1}

Nanoelectronics is currently an exciting field full of possibilities \cite{Akinwande2014,RevModPhys.84.119}. The recent development of materials science and engineering allowed for the description of several interesting materials \cite{doi:10.1002/pssb.201046583,Hirsch2010,doi:10.1002/jctb.1693}. Such a description made it possible to conceive materials with different properties which, in turn, give rise to electronic devices with superior performances \cite{Zapata2016,DIAO2017359,doi:10.1021/nn101671t}. Graphene stands out as the most studied and applied between these materials \cite{lu_JMCC,geim,Novoselov666,Novoselov,doi:10.1021/cr900070d}. Because graphene sheets lack the bandgap characteristic of semiconductors, the electronic industry is particularly interested in studying Armchair Graphene Nanoribbons (AGRNs), i.e., narrow strips of graphene whose lateral edges present a specific kind of symmetry \cite{Kimouche2015}.

As in other organic systems, it is known that a quasi-particle mediated transport takes place in AGNR \cite{neto_JPCL,cunha_PRB,fischer_CARBON,cunha_CARBON}. This is due to the balance of energy presented between the electronic and the lattice degrees of freedom of the system. Thus, depending on the degree of coupling between electrons and lattice, the system presents different transport properties \cite{ribeiro_JPCL,pereirajr_2017a,pereirajr_jmcc_2020,abreu_jmm,pereirajr_sm,pereirajr_jmm_2019a,pereirajr_2019_jpcc,pereirajr_2020_pccp}. A conclusion that directly follows from these considerations is that, to obtain a correct description of the electronic devices based on these materials, it is also crucial to understand the properties of its charge carriers, particularly when under the action of an external electric field.

One of the most common types of charge carriers that take place in an extended organic system are bipolarons \cite{C3NJ00602F,bredas_ACR}. Such structures usually arise when the density of polarons are high \cite{Heeger2001}. In this case, it is usual that the two equally charged polarons might lower their energies by sharing the same distortion. Therefore, bipolarons consist of the coupling state between two equally charged polarons, thus presenting a spinless structure with $\pm 2e$ charge \cite{Heeger2001}.

In this work, we investigate the properties of bipolarons in a narrow AGNR (4-AGNR, in which four carbon atoms define its width). We investigate the behavior of these charge carriers under different electric field and electron-phonon coupling regimes. As an important result, we were able to describe how the effective mass and the terminal velocities of these quasiparticles are impacted by the inclusion of external excitation. 

\section{Methodology}
\label{sec2}

We make use of a 2D version of the SSH-type Hamiltonian \cite{su_PRB,su_PRL,cunha_PRB}, which is analogous to a tight-binding with lattice relaxation in a first-order expansion. The electronic degrees of freedom of the system are described in a second quantization formalism and the lattice is treated classically \cite{su_PRL}. Because the two realms are coupled, however, the problem ought to be solved self-consistently \cite{su_PRB}.

\begin{figure}[pos=ht]
	\centering
	\includegraphics[width=0.4\linewidth]{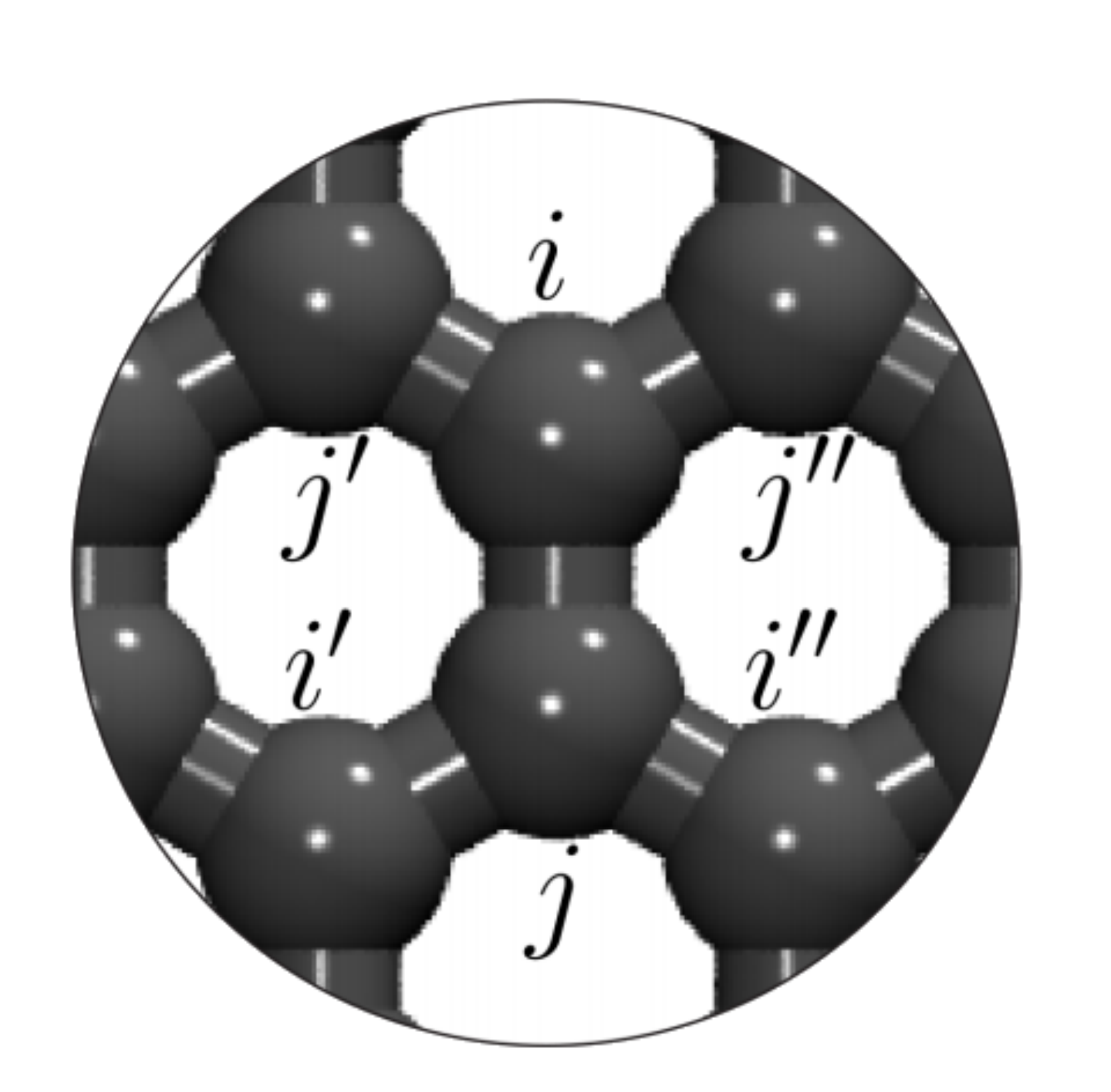}
	\caption{Schematic representation of the labeling system adopted here for the AGNR sites.}
	\label{fig:lattice}
\end{figure}

The electronic transfer integral expresses the hopping of electrons between sites and is dependent on the lattice structure as follows:
\begin{equation}
	\label{hop}
	t_{i,j} = t_0 - \alpha \eta_{i,j}.
\end{equation}
Fig. \ref{fig:lattice} presents the labeling of the sites. Here, $\eta_{i,j}$ is the variation in bond-lengths between two neighboring sites $i$ and $j$. $t_0$ is the hopping integral of the evenly displaced system and $\alpha$ is the electron-phonon coupling constant. Note that the linear dependence of the hopping integral with the site's position variation is justified because such displacement is a small fraction of the equilibrium size (usually smaller than 2\%) \cite{da2016polaron}.

The model semi-classical Hamiltonian of the system is given by:
\begin{equation}
	H = -\sum_{\langle i,j \rangle, s} \left(t_{i,j}^{\phantom{*}} C_{i,s}^\dag C_{j,s}^{\phantom{\dag}}+t_{i,j}^* C_{j,s}^\dag C_{i,s}^{\phantom{\dag}} \right )+ \frac12K\sum_{\langle i,j \rangle} \eta_{i,j}^2+\frac1{2M}\sum_ip_i^2.
	\label{hamiltonian}
\end{equation}
In Equation \ref{hamiltonian}, $ \langle i,j \rangle $ dictates that the sum is to be carried out on over sites only \cite{su_PRB}. Here $C_{i,s}^{\phantom{\dag}}$ is the $\pi$-electron annihilation operator on site $i$ with spin $s$ and $ C_{i,s}^{\dag} $ its hermitian conjugate, i.e., the corresponding creation operator. $\displaystyle\frac12K\sum_{\langle i,j \rangle} \eta_{i,j}^2$ represents the effective potential associated with $\sigma$-bonds between carbon atoms in a harmonic approximation with $K$ being the elastic constant. $\displaystyle\frac1{2M}\sum_ip_i^2$ is the kinetic energy of the sites, $ p_i $ and $M$ are the momenta and mass of the carbon atoms. The Hamiltonian parameters adopted here were previous reported in the literature: 2.7 eV for $t_0$ and 21 eV/\AA$^2$ for $K$ \cite{barone2006electronic,de2012electron,ribeiro2015transport, kotov2012electron,yan2007electric,neto2009c,yan2007raman,cunha_PRB,cunha_CARBON,fischer_CARBON,silva_SR}. The values for the electron-phonon coupling constant, are studied here and discussed on the results section. 

Beginning with a initial state, for the lattice part of the problem, we obtain the expectation value of the system's Lagrangean, $ \langle L \rangle = \langle \Psi | L | \Psi \rangle $, where $ | \Psi \rangle $ is the Slater determinant, and
\begin{equation}
	\langle L \rangle  = \frac{M}{2}\sum_i \dot\xi_i^2-\frac12K\sum_{\langle i,j \rangle}\eta_{i,j}^2 + \sum_{\langle i,j \rangle, s} \left( t_0-\alpha \eta_{i,j} \right)\left(B_{i,j}+B^*_{i,j}\right)
\end{equation}
where,
\begin{equation}
	B_{i,j} \equiv \sum_{k,s}{'}\psi^*_{k,s}(i) \psi^{\phantom{*}}_{k,s}(j).
\end{equation}
This term couples the electronic and the lattice degrees of freedom of the system.  

Following, we can solve the Euler-Lagrange equation to obtain an initial set of coordinates $\{ \eta_{i,j} \}$. Considering such set, a corresponding electronic set $\{ \psi_{k,s}(i) \}$ is obtained and a new Lagrangean is used to return another set of coordinates $\{ \eta_{i,j} \}$. This process is repeated until convergence is achieved, thus obtaining a stationary solution $\{ \eta_{i,j} \}$ and $\{\psi_{k,s}(i)\}$. 

The time evolution of the system is carried out by considering the time-dependent Schr\"{o}dinger equation for electrons:
\begin{equation}
	|\psi_{k,s}(t+dt)\rangle = e^{-\frac i\hbar H(t) dt}|\psi_{k,s}(t)\rangle.
\end{equation}
The procedure to solve such an equation is detailed in previous works \cite{lima2006dynamical,cunha_PRB}.

The dynamics of the lattice part is obtained by the Euler-Lagrange equations, which results in the following Newtonian equation
\begin{equation}
	M\ddot \eta_{i,j} = \frac12K\left(\eta_{i,j'}+\eta_{i,j''}+\eta_{j,i'}+\eta_{j,i''} \right )-2K\eta_{i,j}
	+ \frac12\alpha\left(B_{i,j'}+B_{i,j''}+B_{j,i'}+B_{j,i''}-4B_{i,j} + \mathrm{c.c.}\right ).
\end{equation}

The external electric field $\mathrm{\textbf{E}}(t)$ is included by considering the following Peierls substitution on the electronic transfer integrals of the system
\begin{equation}
	t_{i,j} = e^{-i\gamma\mathrm{A(t)}}\left(t_0 - \alpha \eta_{i,j} \right ).
\end{equation}
$\mathrm{A}(t)=\mathrm{\textbf{A}}(t)\cdot \hat{\eta}_{i,j}$; $\mathrm{\textbf{A}}(t)$ is the vector potential whose relation to the electric field $\mathrm{\textbf{E}}(t)$ is $ \mathrm{\textbf{E}}(t) = -(1/c)\dot{\mathrm{\textbf{A}}}(t) $. $ \gamma \equiv ea/(\hbar c) $, $a$ is the lattice parameter ($ a = 1.42 $ \AA~ in graphene nanoribbons), $e$ is the absolute value of the electronic charge, and $c$ the speed of light. In this work, we turned the electric field on adiabatically \cite{da2016polaron}.

\section{Results}
\label{sec3}

\begin{figure}
	\centering
	\includegraphics[height=9cm]{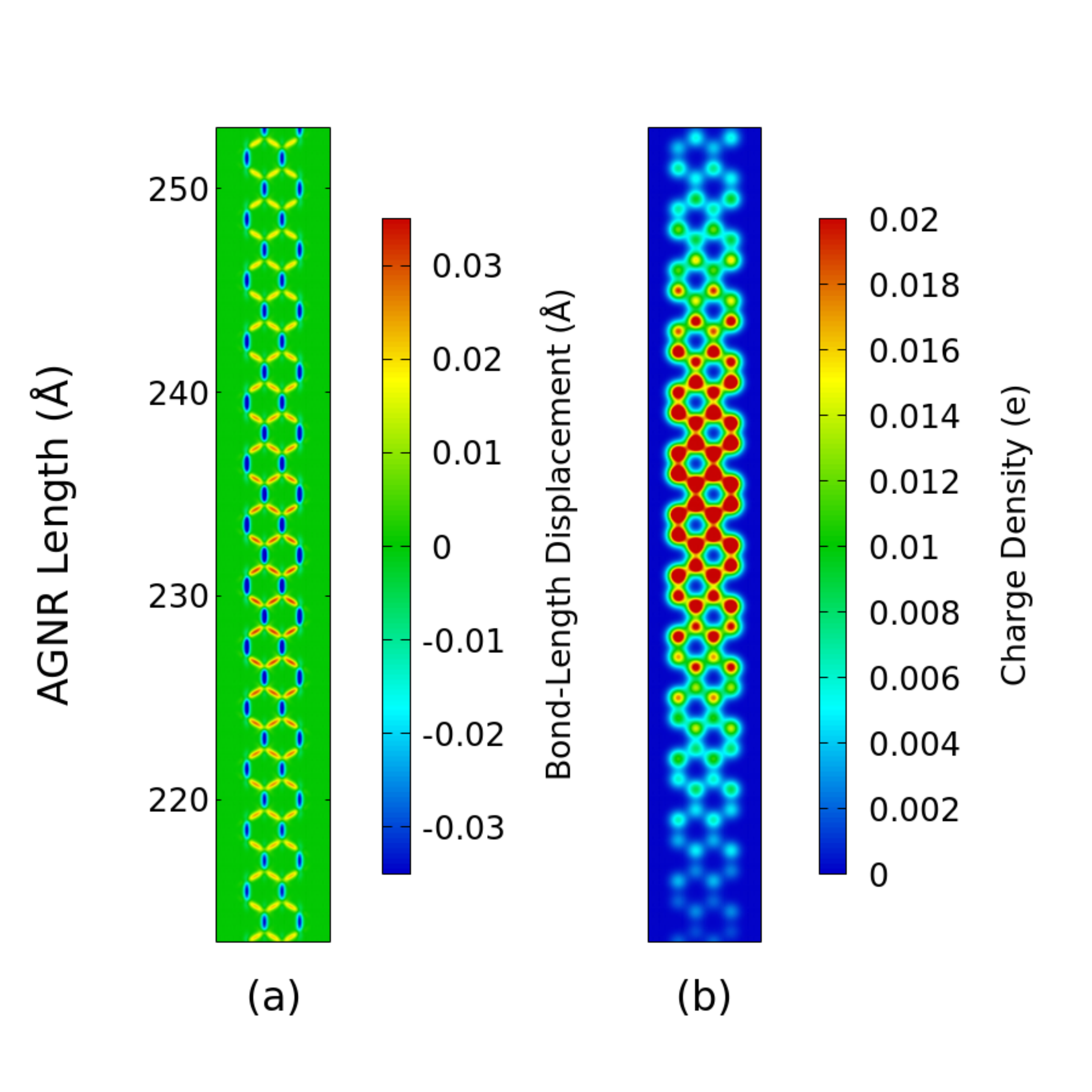}
	\caption{(a) The Bond-length pattern in 4-AGNR with containing a bipolaron; (b) the respective charge density profile.}
	\label{Fig.2}
\end{figure}

The main propose of this work is to investigate how the bipolaron dynamic is influenced under different intensities of electric field and electron-phonon coupling constant in a narrow AGNR, i.e., one with four atoms width. By performing dynamics simulations, this kind of investigation allows one to compute the effective mass of this quasiparticle, which is crucial to the description of AGNRs as candidates to develop electronic devices. We begin by presenting, in Fig. \ref{Fig.2}, the stationary properties of a bipolaron in AGNR. Fig. \ref{Fig.2}(a) presents the bond-lengths pattern of the bipolaron endowed system and Fig. \ref{Fig.2}(b) the respective charge density. The direct relation between charge localization and related lattice deformations are evidence of the quasiparticle formation in the organic materials\cite{Heeger2001,pereirajr_jpcl,pereirajr_jpcc_2019b}. As we showed in Fig. \ref{Fig.2}(b) the bipolaron's charge extended for approximately $30$ \r{A}. Fig. \ref{Fig.2}(a) shows that this is the same region in which a more pronounced displacement of the bond-length is present.

\begin{figure}
	\centering
	\includegraphics[height=12cm]{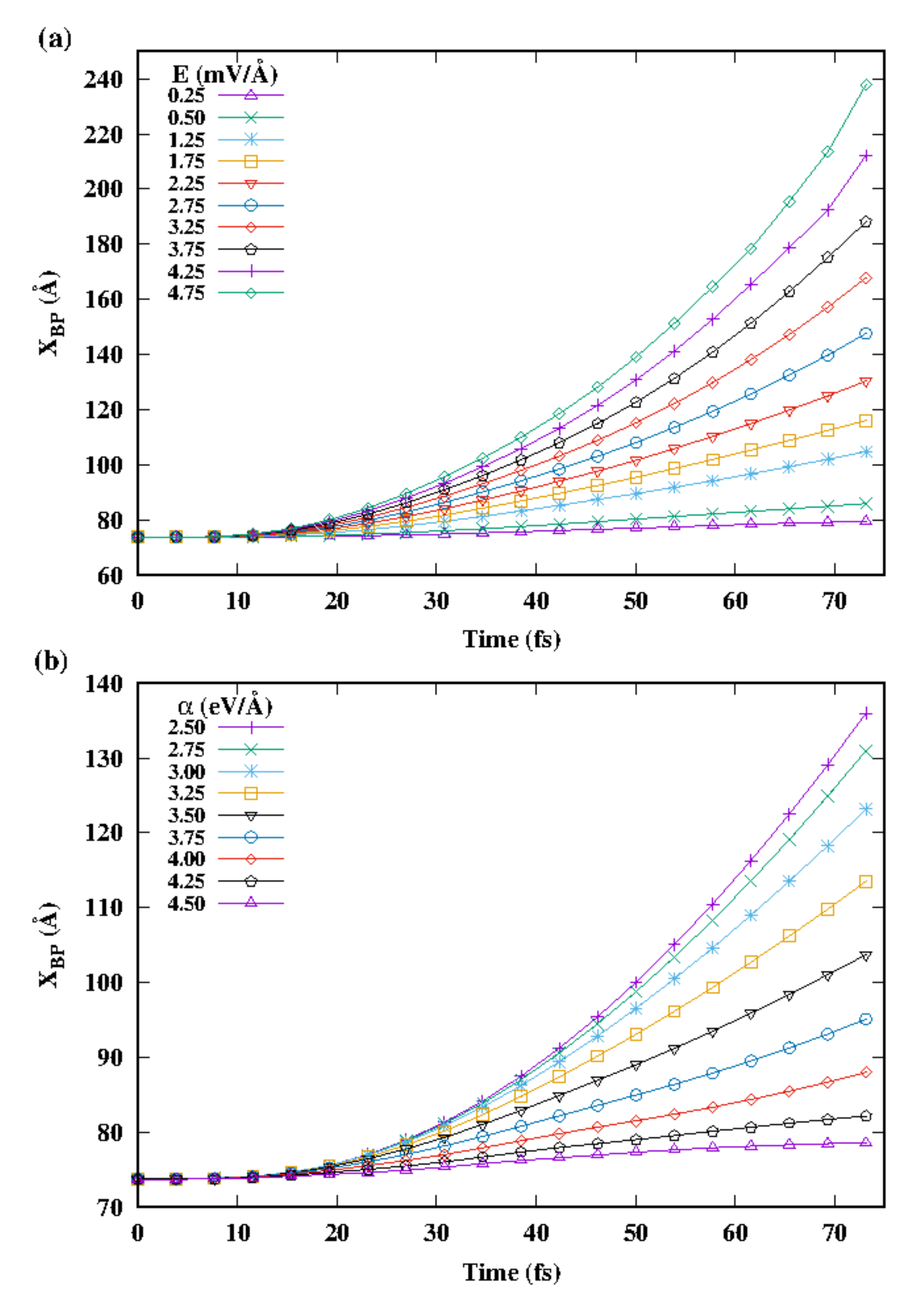}
	\caption{(a) Dynamics a bipolaron for different electric field intensities; (b) dynamics of a bipolaron in for different electron-phonon coupling constant.}
	\label{Fig.3}
\end{figure}

We proceed by investing how the different intensities of external electric fields influence the bipolaron motion through the center of charge movement \cite{ribeiro_NJC}. For this purpose, we maintain the electron-phonon coupling at the constant value of $3.2$ eV/\r{A}. This is a low value when compared to what is normally used \cite{Cunha2014,Lima2006} and was adopted so that the stability of the quasiparticle could be verified as a sole effect of the different electric fields intensities. These results are shown in Fig. \ref{Fig.3}(a) to observe the variation of the average position of the center of charge ($X_{BP}$) as a function of time for various electric field intensities. Although periodic boundary conditions were applied to the system, the time interval considered was chosen so that, in all cases considered, the quasiparticle does not reach the edge of the nanoribbon. The results suggest that the greater the intensity of the electric field the more quickly the quasiparticle travel through the nanoribbon length, which is an expected feature from previous works in the literature\cite{Feng2014,Schwierz2010}. The fact that the curves are rather smooth is indicative of the stability of the bipolaron. This is true even for the largest considered electric fields. A larger dispersion throughout the nanoribbon would lead to imprecise results in relation to the position of the charge center. This could reflect the loss of stability of the quasiparticle, as charge would tend to spread over the entire length of the nanoribbon, thus being dissociated from the lattice distortion that characterizes it. 

In Fig. \ref{Fig.3}(b) we present the results of the bipolaron dynamics for different values of the electron-phonon coupling constant. In these cases, the electric field intensity was maintained constant in $1.0$ mV/\r{A}. Again, the results refer to the movement of the charge center where the time interval was chosen so that the quasiparticle would not reach the edge of the system. The results show that, as the electric field, the electron-phonon coupling constant has a significant influence in the quasiparticle mobility. The higher the $\alpha$ values, the more slowly the bipolaron moves along the nanoribbon length. These results also suggest that for the two largest coupling constants considered, the bipolaron movement is approximately $60$ fs delayed when compared to the smallest one. These results are consistent to the fact that the electron-lattice coupling constant is related to how the lattice trapped the quasiparticle, as described by our model. Therefore a strong influence in the movement of the quasiparticle is indeed expected. 

The simulations that gave rise to Fig. \ref{Fig.3} readily allows one to compute how the average velocity of this quasiparticle is influenced by the electric field. The results of these calculations are shown in Fig. \ref{Fig.3}(a), from which it can be seen that the interplay between field strength and terminal velocity is nearly linear. With such a description in hand, we can now investigate the relationship between the effective mass of the bipolaron in the system ($m_{eff}$) and the electron-phonon coupling as well as to the electric field. For the calculation of the effective mass, we considered the same scheme as employed for the calculation of the effective mass of the polaron and bipolaron in cis-polyacetylene and poly-p-phenylene (PPP) \cite{Falleiros2019}, i.e., the equality between impulse and linear momentum variation ($\Delta \vec{p}= q\vec{E} \Delta t)$. The electric field was turned on adiabatically so that to avoid artificial numerical effects \cite{Cunha2016}. Our analysis focused on the time interval necessary for the quasiparticle to reach the saturation velocity. For the calculation of the linear momentum variation, we took the velocity of the bipolaron at the end of the time interval considered for each case shown in Fig. \ref{Fig.3}.

\begin{figure}
	\centering
	\includegraphics[height=15cm]{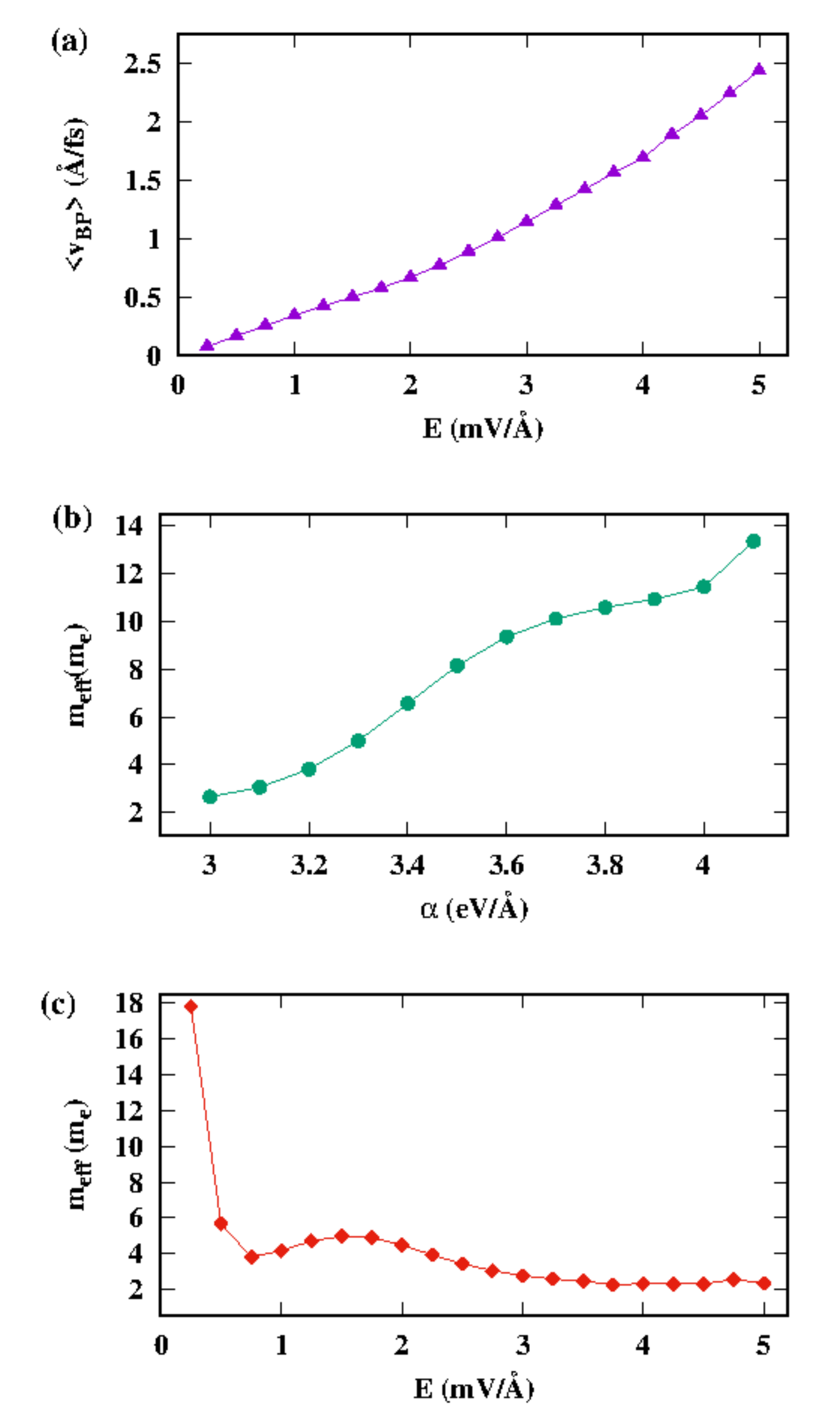}
	\caption{(a) The average velocity of the bipolaron in 4-AGNR with different electric field intensities; (b) effective mass of the bipolaron as a function of the electron-lattice coupling constant; (c) effective mass of the bipolaron as a function of the electric field strength.}
	\label{Fig.4}
\end{figure}

The results of the different values of the electron-phonon coupling constant and the respective variation of effective mass are shown in Fig. \ref{Fig.4}(b). For these cases, the electric field intensity was set to be $1.0$ mV/\r{A}. As can be seen from Fig. \ref{Fig.4}(b), the effective mass of the bipolaron increases rapidly with an increment of the electron-phonon coupling constant. Considering $m_e$ to be the rest mass of a free electron, for $\alpha = 3.0$ eV/\r{A} the value calculated for the effective mass of the bipolaron is $2.6$ $m_e$. It increases to $13.4$ $m_e$ for $\alpha = 4.1$ eV/\r{A}. For $\alpha$ values between $3.6$ eV/\r{A} and $4.0$ eV/\r{A} the increase in the effective mass presents a slower increase, thus resulting in a practically linear behavior.

Finally, in Fig. \ref{Fig.4}(c) we show the relation between the effective mass and the electric field intensity. In these cases, the electron-phonon coupling constant is kept as $3.2$ eV/\r{A}. For $E=0.25$ mV/\r{A} the effective mass calculated for a bipolaron was $17.8$ $m_e$ and for the next value considered, it dramatically reduces to $5.7$ $m_e$. Between $0.75$ mV/\r{A} and $1.5$ mV/\r{A} the value of the effective mass increases again until $5.0$ $m_e$. Following, it presents a subtle decrease. For electric field intensities higher than $3.75$ mV/\r{A} the effective mass is practically constant around $2.2$ $m_e$. These results allow us to verify that the electric field has a significant influence on the effective mass of the bipolaron in AGNRs.

\section{Conclusions}
\label{sec4}

In summary, the dynamics of bipolaron in a narrow armchair graphene nanoribbon was numerically studied in the framework of a 2D tight-binding Hamiltonian in which electron-phonon interactions were taken into account. The dynamics of the quasiparticle was evaluated as a function of the electric field strength and electron-phonon coupling constant, two relevant factors for charge carriers dynamics. Our results have shown that bipolarons are stable quasiparticle even in a regime of high electron field strength. Furthermore, the increase in electron-phonon coupling constant leads to a reduction in its mobility. We were also able to calculate the effective mass and terminal velocities of the bipolarons and their relation to the aforementioned properties. Our results have shown that the effective mass decreases with increasing electric field strength and increases with an increasing electron-phonon coupling constant.

\section*{Acknowledgements}
The authors gratefully acknowledge the financial support from Brazilian Research Councils CNPq, CAPES, and FAPDF and CENAPAD-SP for providing the computational facilities. L.A.R.J. gratefully acknowledges the financial support from FAPDF grant 00193.0000248/2019-32 and CNPq grant 302236/2018-0.

\printcredits
\bibliographystyle{unsrt}
\bibliography{cas-refs}

\end{document}